\documentclass{ws-ijmpd}
\begin{document}

\markboth{A.A. Ivanov and L.V. Timofeev}
{Temporal signatures of the Cherenkov light induced by EAS}

\title{Temporal signatures of the Cherenkov light induced by extensive air showers of cosmic rays detected with the Yakutsk array}

\author{A.A. Ivanov* and L.V. Timofeev}

\address{Shafer Institute for Cosmophysical Research and Aeronomy,\\
Lenin Avenue 31, Yakutsk 677980, Russia\\
*ivanov@ikfia.ysn.ru}

\maketitle

\begin{abstract}
We analyze temporal characteristics of signals from the wide field-of-view (WFOV) Cherenkov telescope (CT) detecting extensive air showers (EAS) of cosmic rays (CR) in coincidence with surface detectors of the Yakutsk array. Our aim is to reveal causal relationships between measured characteristics and physical properties of EAS.
\end{abstract}


%%%%%%%%%%%%%%%%%%%%%%%%%%%%%%%%%%%%%%%%%%%%%%%%%%%%%%%%%%%%%%%%%%%%

\section{Introduction}	

Galactic cosmic rays are investigated by extensive air shower arrays, among other approaches, for many years. Some interesting features of the energy spectrum and mass composition of CRs were elucidated, but other challenging questions remain unsolved\cite{Amato}. For example, where are the sources of galactic CRs? Are they supernova remnants or other peculiar objects? Is the 'knee' of CR spectrum due to the diffusion of particles in magnetic fields or to the upper limit of galactic sources? Where in the energy scale is the transition region between galactic and extragalactic components of CRs?

One possible way to strengthen EAS array capabilities is to equip it with Cherenkov telescopes detecting EAS of CRs in coincidence with surface detectors.\cite{Jelley,Chudakov} Coherent Cherenkov radiation of EAS particles produces more intensive light spot on the ground than the isotropic atmospheric nitrogen fluorescence light induced by the passage of the shower, so the small and cheap telescopes can be used to detect the signal.

A subset of WFOV Cherenkov telescopes is able to measure the depth of cascade maximum, $x_m$, and/or the shower age via angular and temporal distributions of the Cherenkov signal. Combining $x_m$ and the shower age with other EAS characteristics measured with surface detectors of the array, e.g. the energy and muon content, one is able to estimate the mass composition of CRs. Experimental arguments in elucidating the origin of the knee and ankle in CR spectrum will significantly strengthen due to the measurements of the angular and temporal distributions of the Cherenkov signal in the energy range above $10^{14}$ eV

\begin{figure}[t]\centering
\includegraphics[width=0.7\columnwidth]{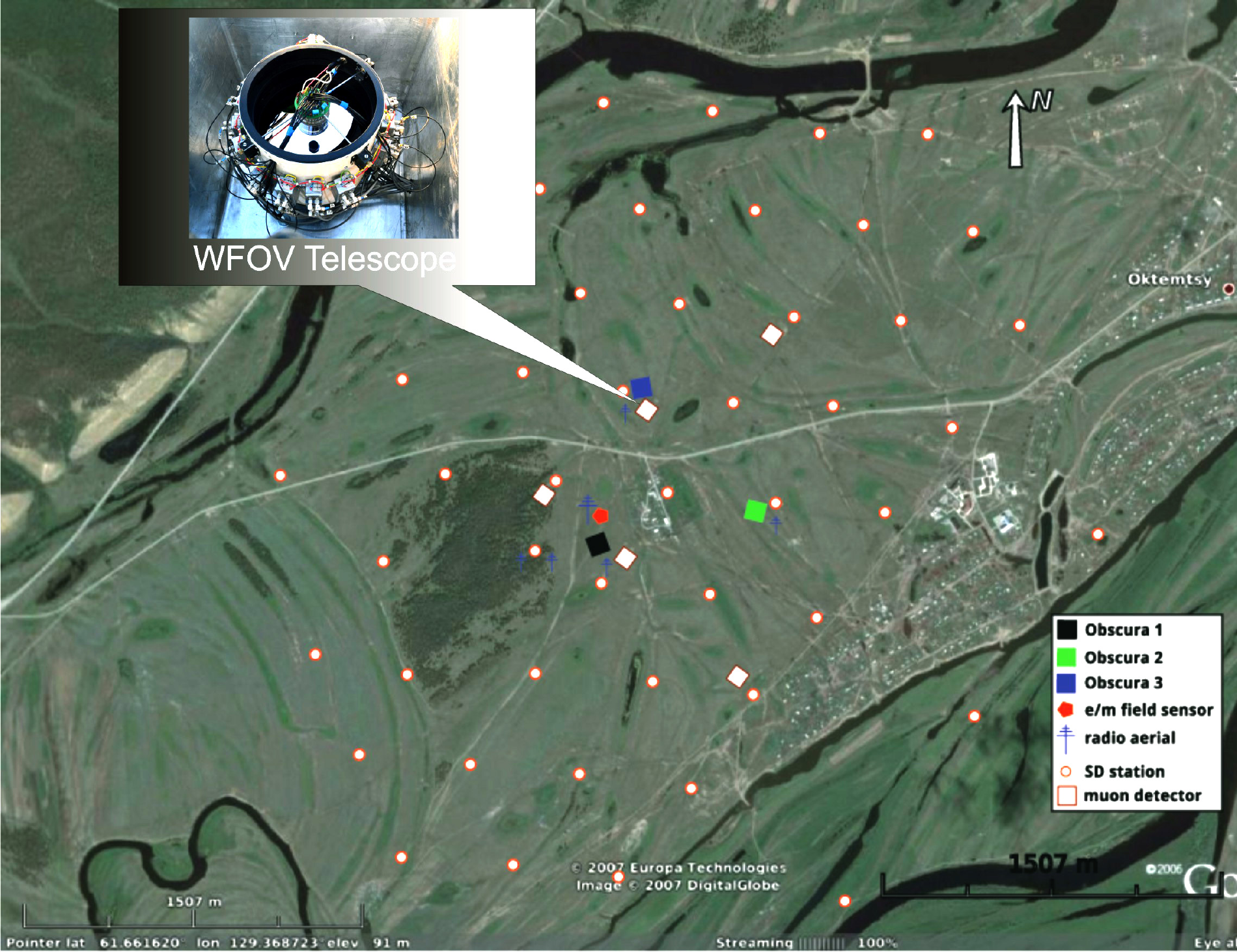}
  \caption{Arrangement of the Yakutsk array detectors. WFOV Cherenkov telescope is placed $\sim500$ m from the array center.}
\label{Fig:Map}\end{figure}

We have constructed a prototype Cherenkov telescope to operate in cooperation with the surface detectors of the Yakutsk array. In the paper, the measured set of temporal characteristics of the Cherenkov light induced by EAS is analyzed in order to elucidate the shower development parameters in the atmosphere.

%%%%%%%%%%%%%%%%%%%%%%%%%%%%%%%%%%%%%%%%%%%%%%%%%%%%%%%%%%%%%%%%%%%%%%%%

\section{A brief description of the apparatus and data acquisition}
The Yakutsk array is located at geographical coordinates $61.7^0$N, $129.4^0$E, at a mean altitude of 105 m above sea level.
The array is measuring EAS with 58 plastic scintillation counters in surface stations, muons with 4 underground detectors, and Cherenkov light with 48 photomultiplier tubes (PMTs). All detectors are distributed within the 8 km$^2$ array area (Fig. \ref{Fig:Map}). The energy range of EAS investigations is $10^{15}$ eV to $10^{19}$ eV.

EAS events are selected from the background using a two-level trigger of detector signals: The first
level is a coincidence of signals from two scintillation counters in a station within $2 \mu s$; the second level is a
coincidence of signals from at least three nearby stations (not lined up) within $40 \mu s$.

The shower core coordinates are located fitting the lateral distribution of particle densities by the
Greisen-type function. Core location errors are $\sim30$ m to $\sim50$ m depending on CR energy.
Arrival angles of the EAS primary particles are calculated in the plane shower front approximation using
detection times at the stations. A clock pulse transmitter at the center of the array provides pulse timing to
$100$ ns accuracy. Errors in arrival angles depend on the primary energy decreasing from $7^0$ at E = 1 EeV ($=10^{18}$ eV)
to $3^0$ above E = 10 EeV. More experimental details and physics results are given in Refs. \refcite{Kashiwa}$-$\refcite{IJMPD}.

The WFOV Cherenkov telescope consists of the spherical mirror ($\varnothing260$ mm, $F=113$ mm) mounted at the bottom of tube and a multi-anode PMT Hamamatsu R2486 ($\varnothing50$ mm) in the focus. A voltage-divider circuit and signal cables are attached to the bearing plate. 32 operational amplifiers are mounted onto the tube. The Cherenkov telescope is mounted vertically near the Obscura 3 station (Fig. \ref{Fig:Map}). A comprehensive description of WFOV CT can be found in Refs. \refcite{ASTRA} -\refcite{NIM}.

\begin{figure}[t]\centering
\includegraphics[width=0.7\columnwidth]{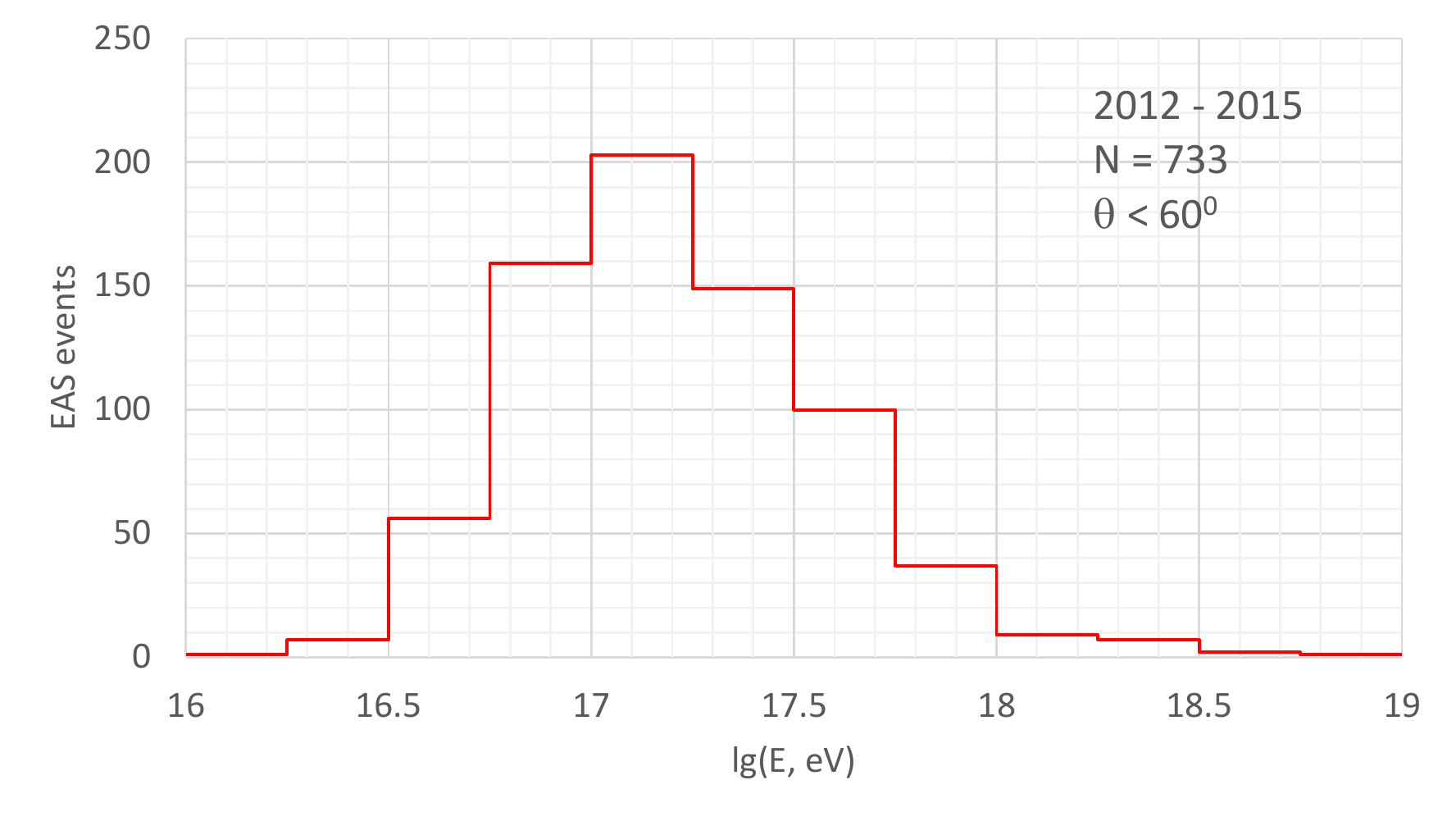}
  \caption{Energy distribution of EAS events detected with the surface detectors of the Yakutsk array and WFOW Cherenkov telescope in coincidence.}
\label{Fig:Energy}\end{figure}

The data are accumulated within the observation period 2012--2015 where coincident EAS events are detected. A sample of showers is selected in the energy interval $(2\times10^{16},10^{19})$ eV with the mean energy $0.3$ EeV. Additional cuts are applied to exclude showers with axes out of the array area and zenith angles $\theta>60^0$. A specific cut is used for saturated signals where EAS events generate more Cherenkov light than CT can detect properly. These signals are rejected, too. A number of EAS events survived after cuts is 733 with the primary energy distribution shown in Fig. \ref{Fig:Energy}.

\begin{figure}[b]\centering
\includegraphics[width=\columnwidth]{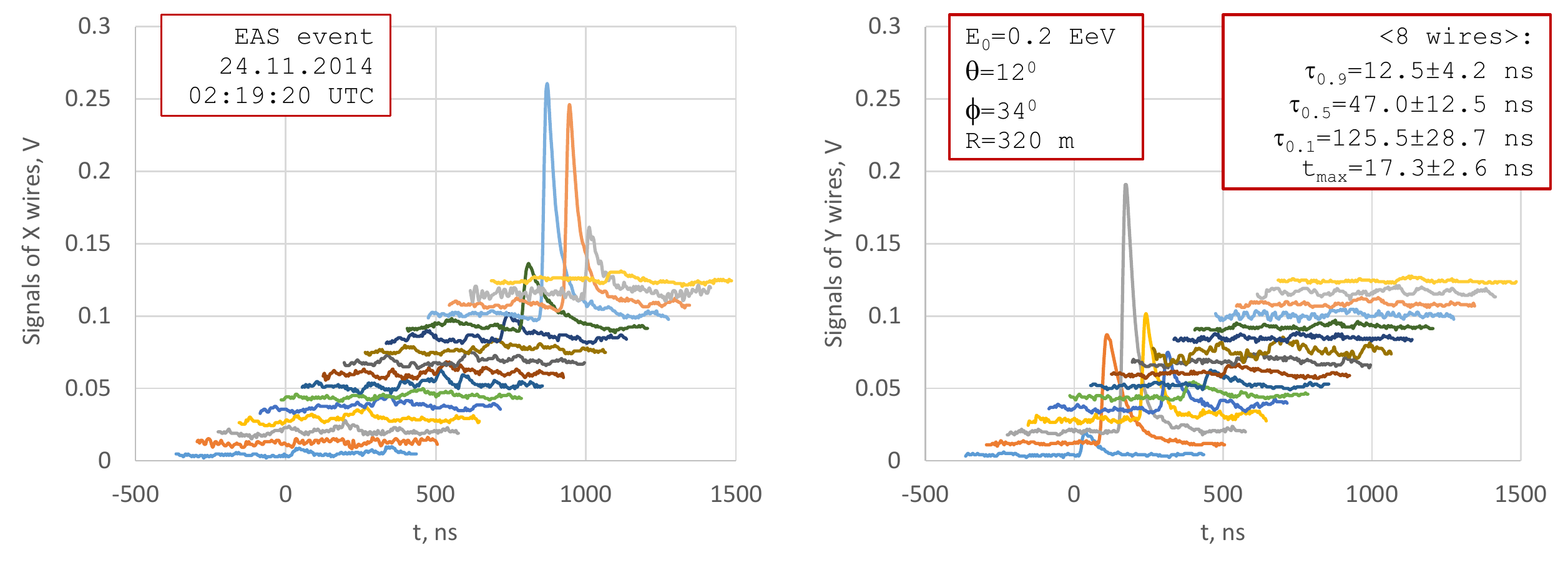}
  \caption{Signals of 32 wires from multi-anode PMT in the focus of telescope. EAS parameters in a particular shower are estimated using data of the surface detectors. Pulse durations ($\tau$) are measured at 0.1, 0.5, 0.9 levels of the pulse maximum $V_{max}$. Time difference ($t_{max}$) between $V_{max}$ and $0.1V_{max}$ of the rising signal is shown in the right panel.}
\label{Fig:Signals}\end{figure}

An example of the set of WFOV Cherenkov telescope output signals in a particular EAS event is shown in Fig. \ref{Fig:Signals}. The average number of wires in events is $17\pm6$ where the signal definitely exceeds the noise.

%%%%%%%%%%%%%%%%%%%%%%%%%%%%%%%%%%%%%%%%%%%%%%%%%%%%%%%%%%%%%%%%%%%%%%%%

\section{Pulse duration as a function of the distance to the shower core}
\begin{figure}[t]\centering
\includegraphics[width=0.7\columnwidth]{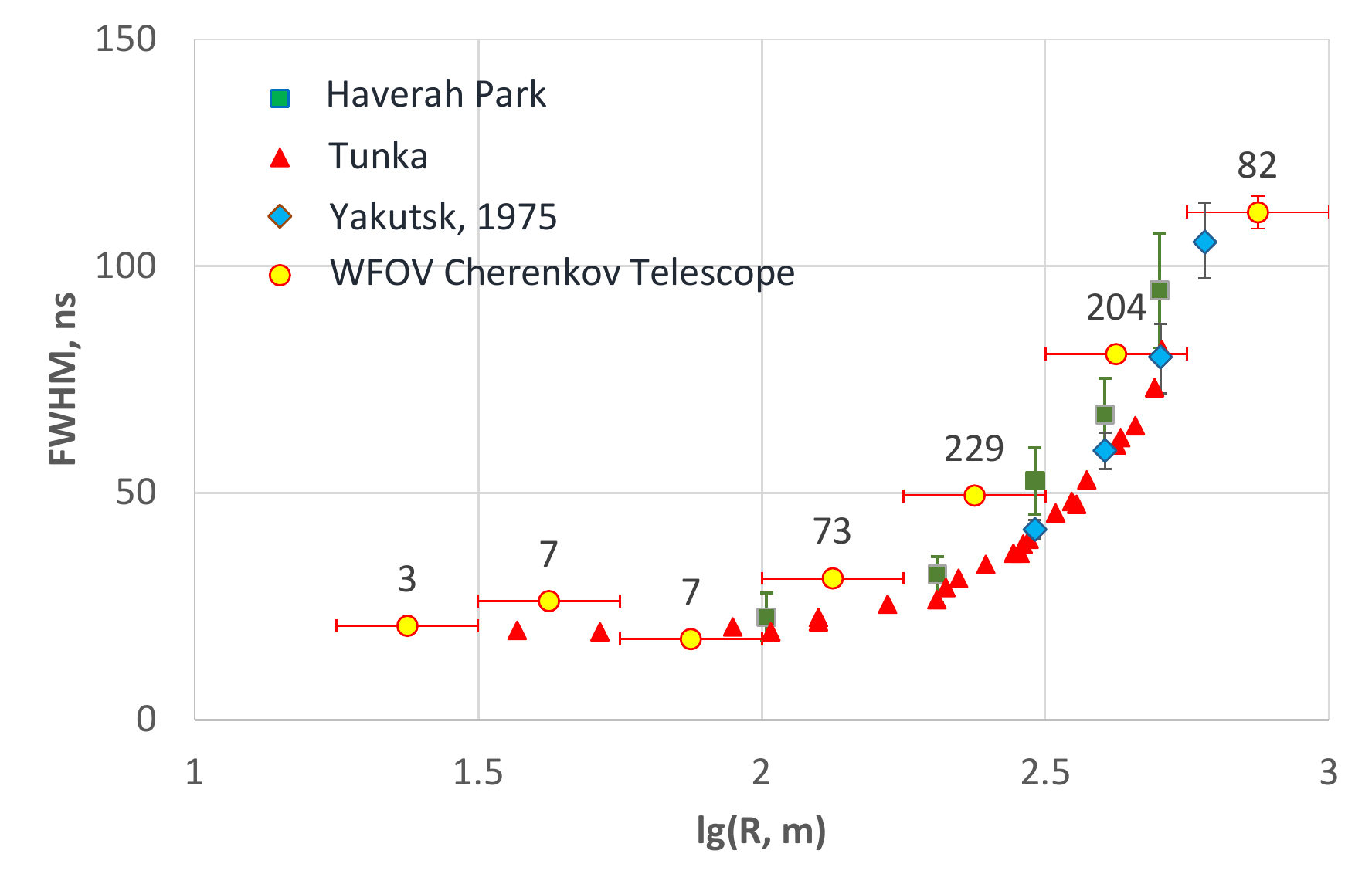}
  \caption{Full width at half-maximum of the Cherenkov signal as a function of the shower core distance, $R$, measured by EAS arrays. HP and Yakutsk data are averaged over samples of showers, while Tunka signal is detected in a particular EAS event. Vertical bars are statistical errors while horizontal bars indicate the shower core distance bins. Event numbers are indicated above our data points.}
\label{Fig:FWHM}\end{figure}

We focus here on the pulse shape parameters of the Cherenkov signal from EAS. The main feature of the signal is its duration rising with the shower core distance due to geometrical reasons. The first measurements of the parameter were made in Yakutsk\cite{Klmkv} and Haverah Park\cite{Trvr}. The most recent results are provided by the Tunka array.\cite{Tunka} Our measurement of the full width at half-maximum (FWHM) of the signal confirms the previous results (Fig. \ref{Fig:FWHM}).

The relevant energy ranges are: Haverah Park $\langle E\rangle\sim0.2$ EeV; Yakutsk, 1975 $E\in(0.001,1)$ EeV; Tunka $E\in(0.003,0.03)$ EeV.

% We have analyzed FWHM dependence on other shower parameters: zenith and azimuth angles, and energy. No significant variation exceeding instrumental errors is found. As an illustration, the signal width is shown as a function of energy in Fig. \ref{Fig:FWHMvsE}. A systematic variation of the width is visible in the shower core distance intervals, although it does not exceed experimental errors.

\begin{figure}[t]\centering
\includegraphics[width=0.7\columnwidth]{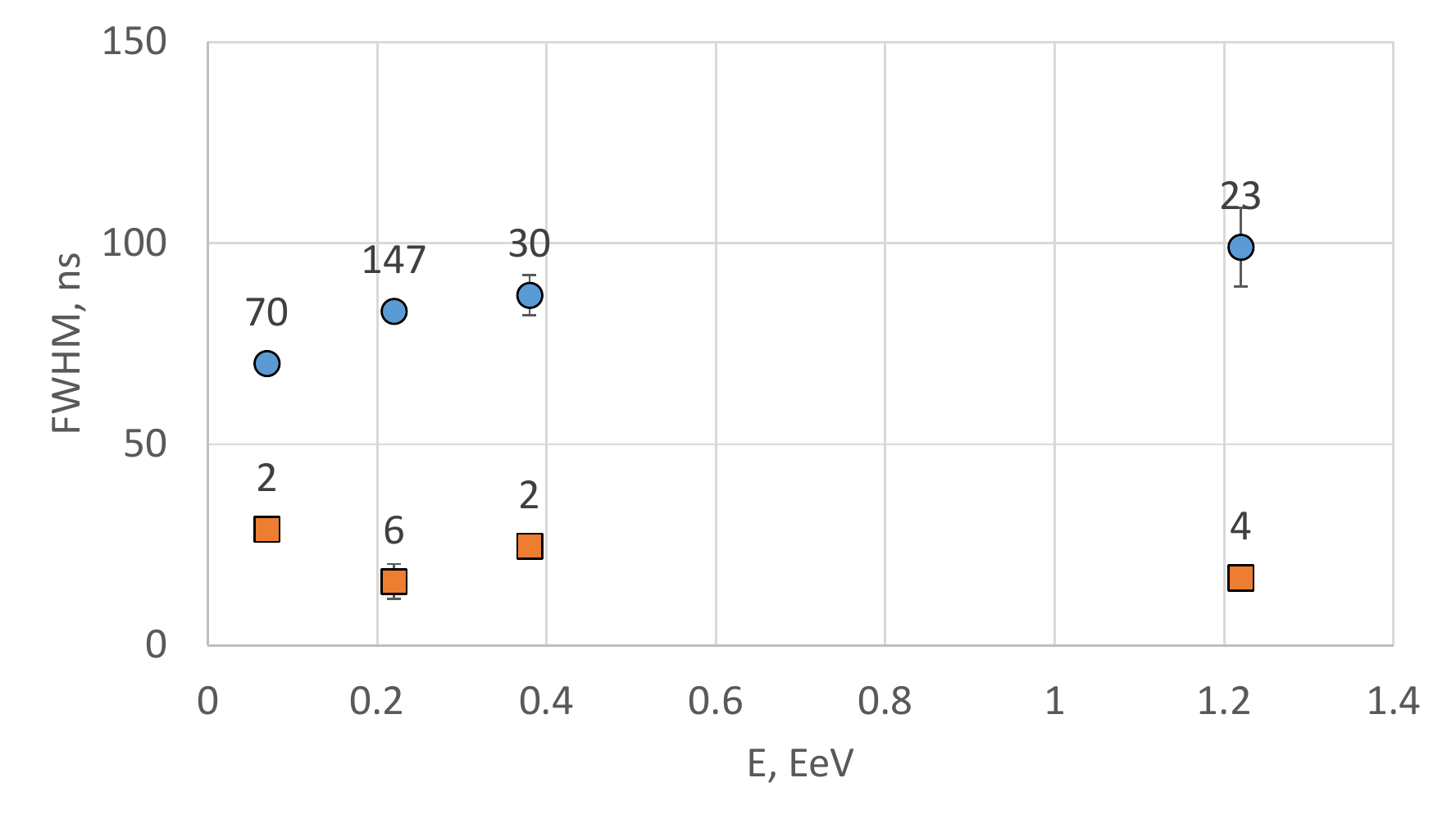}
  \caption{Energy dependence of FWHM. The signal width is measured in EAS core distance intervals $R\in(56,100)$ m (squares) and $R\in(316,562)$ m (circles). Event numbers are indicated above data points.}
\label{Fig:FWHMvsE}\end{figure}

We have analyzed FWHM dependence on other shower parameters: zenith and azimuth angles, and energy. No significant variation vs angles is found exceeding instrumental errors. On the other hand, an indication of the energy dependence of the pulse width is found in our data. As an illustration, the signal width is shown as a function of energy in Fig. \ref{Fig:FWHMvsE}. A systematic variation of the width is visible, although the tendencies are contradictory in the shower core distance intervals. More data are needed, especially at $R<100$ m, to reveal the reliable energy dependence.

Application of the FWHM($R$) function is possible in the shower core location procedure in the array plane. The method was pioneered by Linsley\cite{Lnsl} using charged particle signal width vs $R$. While in the Yakutsk array group, the particle and Cherenkov photon density distribution functions are used to locate the shower core, the FWHM($R$) function can be used additionally (or instead) to refine the core coordinates in the array plane.

%%%%%%%%%%%%%%%%%%%%%%%%%%%%%%%%%%%%%%%%%%%%%%%%%%%%%%%%%%%%%%%%%%%%%%%%

\section{Pythagorean method to estimate the distance to the shower maximum via Cherenkov light measurements}
Common methods in use to find the height in the atmosphere, $h_m$, where the number of EAS particles reaches a maximum, which rely on Cherenkov light detectors, are based on the measurement of the lateral distribution of the photon density and pulse duration.\cite{Ngn}$^-$\cite{Brzk}
We propose another approach to $h_m$ evaluation resulting from our measurements of the Cherenkov light induced by EAS.

Having the time delay, $\Delta t$, of the Cherenkov signal maximum in detector relative to the shower axis crossing the array plane at distance $R$, one is able to calculate the distance, $h_{\theta}$, to the height in atmosphere, $h_m^{Cher}$, where the emission of Cherenkov photons reaches a maximum. Other shower parameters needed are the coordinates of the shower axis and the EAS arrival angles. In our case, these parameters can be provided by the set of synchronized surface detectors of the Yakutsk array.

\begin{figure}[t]\centering
\includegraphics[width=0.7\columnwidth]{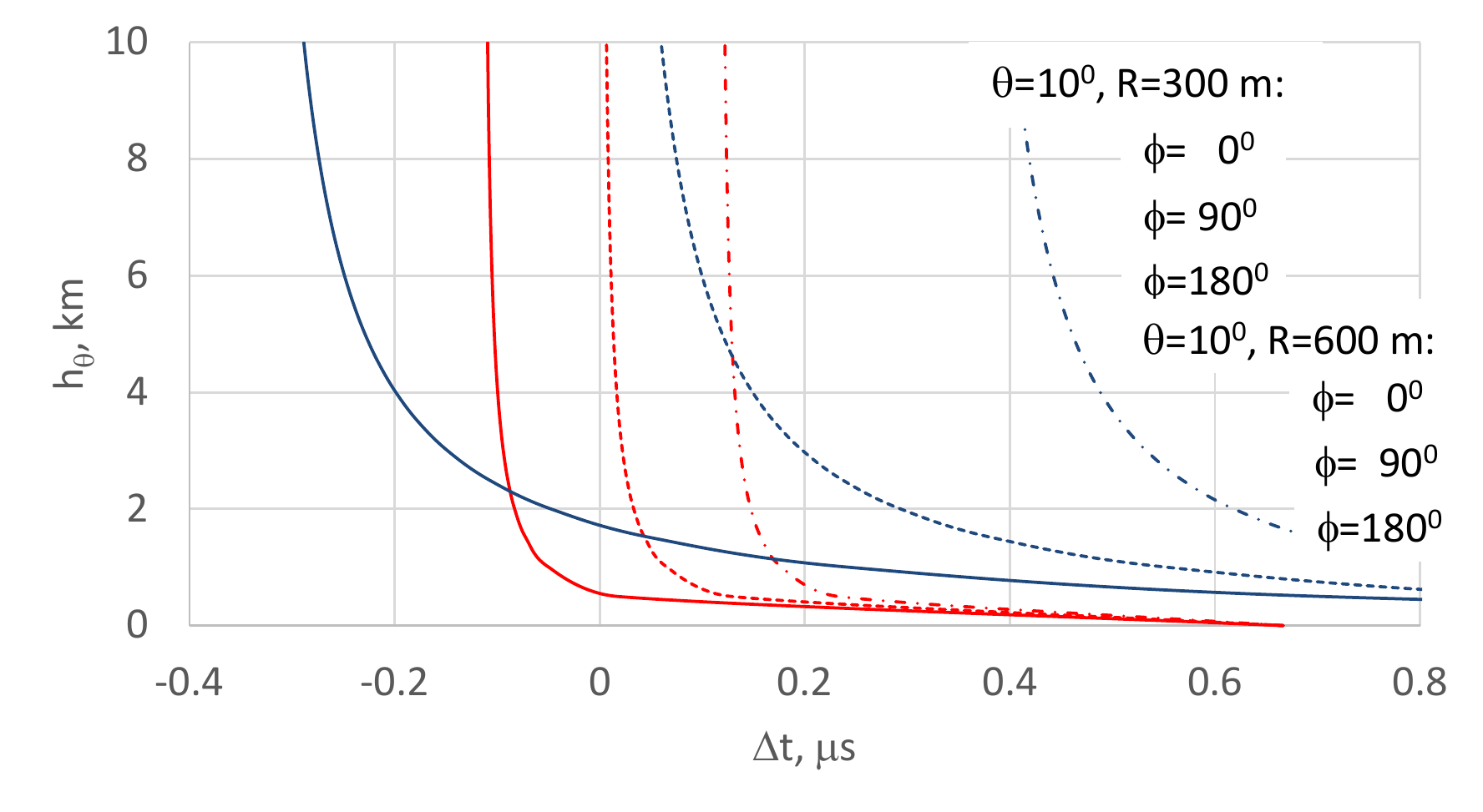}
  \caption{The height of Cherenkov light emission maximum as a function of time difference, $\Delta t$, between Cherenkov photons arriving at the detector and at the shower core in the array plane. Azimuth angle, $\phi$, is between the detector and the shower axis projection on the array plane.}
\label{Fig:Hmax}\end{figure}

The time difference is determined by triangles consisting of $R,h_m^{Cher},h_{\theta}$; that is why we call the method `Pythagorean':
$$
c\Delta t=
\sqrt{h_{\theta}^2+R^2-2Rh_{\theta}\sin\theta\cos\phi}-h_{\theta},
$$
where $\phi,\theta$ are shower arrival angles; c=0.3 m/ns. In a previous paper\cite{NIM} a relation is given with slightly different notation. A solution is given by
$$
h_{\theta}=\frac{0.5(R^2-(c\Delta t)^2)}{c\Delta t+R\sin\theta\cos\phi},
$$
and is illustrated in Fig. \ref{Fig:Hmax}.

Rather large values of the derivative $dh_{\theta}/d\Delta t$ near the time difference threshold restrict the maximum location possibilities within narrow limits due to moderate timing accuracy. On the other hand, relatively small heights can be resolved with reasonable timing resolution.

\begin{figure}[b]\centering
\includegraphics[width=0.7\columnwidth]{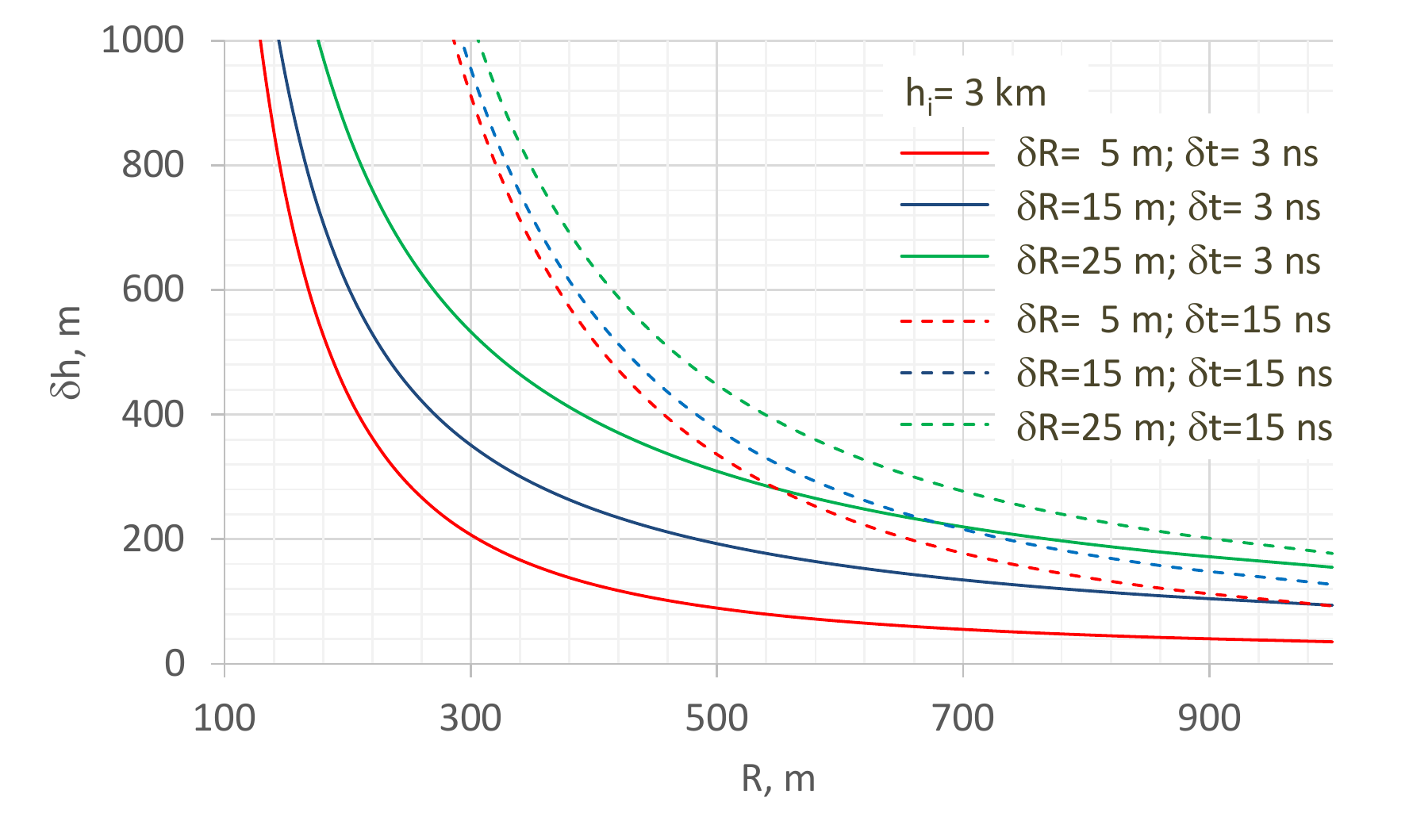}
  \caption{Estimation of the $h_m$ reconstruction accuracy as a function of the shower core distance. Different EAS core location and timing errors are given.}
\label{Fig:Accuracy}\end{figure}

We illustrate in Fig. \ref{Fig:Accuracy} the maximum height location accuracy achievable with the different shower core and time measurement errors.
%----
If the shower core distance is limited to $R>500$ m, then the optimistic estimate is $\delta x_m^{Cher}/x_m^{Cher}<1.3\%$ for $\delta R=5$ m, $\delta t=3$ ns, while the pessimistic estimate is $\delta x_m^{Cher}/x_m^{Cher}<6.5\%$ for $\delta R=25$ m, $\delta t=15$ ns. Here, the isothermal approximation of the atmosphere below $h_0=6.9$ km is used to evaluate the relative error of the depth $\mid\delta x/x\mid=\delta h/h_0$.
%----

The present accuracy for the Yakutsk array detectors (see Section 2) is not sufficient to measure the distance to the shower maximum with reliable results. Significant improvements are needed for the timing resolution and synchronization of detectors. The key objective in this context is improvement of synchronization of the array detectors in order to refine the core location accuracy and measurement of time differences between signals in detectors. The Yakutsk array modernization program\cite{ASTRA,Plans} in progress now includes the best part of the improvements needed.

%---
Particularly, a target synchronization accuracy of detectors is $\delta t=5$ ns to be implemented with Gigabit Ethernet via optical fibers. New scintillation counters and Cherenkov light detectors will provide the core location accuracy of showers within $15-20$ m. Accordingly, the modernized array will be able to measure $x_m^{Cher}$ with accuracy $3-4\%$, if the time resolution of the Cherenkov light detectors will be better than $5$ ns.
%---

While the height of Cherenkov light maximum, $h_m^{Cher}$, is determined by geometrical values only, such as the shower core distance, zenith angle, etc., estimation of the shower maximum\footnote{in the total number of electrons, $h_m$} position is more complicated. The angular and radial distribution of electrons in the shower has an influence on the maximum location. Model simulations of the shower development in the atmosphere and the Cherenkov light emission should be used to reconstruct $h_m$ based on the evaluation of $h_m^{Cher}$.\cite{ASTRA}

%%%%%%%%%%%%%%%%%%%%%%%%%%%%%%%%%%%%%%%%%%%%%%%%%%%%%%%%%%%%%%%%%%%%%%%%
\section{Conclusions}
Coincident detection of EAS events by the surface detectors of the Yakutsk array and WFOV Cherenkov telescope is performed over three years. A total of 733 showers are selected for analysis applying quality cuts.

A FWHM parameter of EAS is measured as a function of the shower core distance. We confirm the results of previous measurements within the primary energy interval $(0.02,10)$ EeV. The function FWHM(R) can be used to locate the shower core in the array plane along with the particle density distribution functions.

A new approach is explored in the estimation of the distance to the shower maximum, based on the time difference between Cherenkov photons arriving at the detector and the shower core in the array plane. Fast optical detectors and nanosecond-order synchronization systems of the array detectors are essential in the future implementation of the method.

%%%%%%%%%%%%%%%%%%%%%%%%%%%%%%%%%%%%%%%%%%%%%%%%%%%%%%%%%
\section*{Acknowledgments}
We are grateful to the Yakutsk array staff for the data acquisition and analysis. The work is supported by RFBR Grant no. 13-02-12036.

\end{document}